\documentclass[twocolumn,showpacs,preprintnumbers,superscriptaddress]{revtex4}
\usepackage{amsmath}
\usepackage{amssymb}
\usepackage{mathrsfs}
\usepackage{graphicx}
\usepackage{bm}
\usepackage{braket}
\usepackage{float}
\usepackage{epstopdf}
\usepackage[colorlinks=true,citecolor=blue,linkcolor=blue,urlcolor=blue]{hyperref}
\hypersetup{colorlinks=true,linkcolor=blue,filecolor=magenta,urlcolor=blue}

\begin{document}

\title{Electronic properties of bilayer phosphorene quantum dots in the presence of perpendicular electric and magnetic fields}

\author{L. L. Li}
\email{longlong.li@uantwerpen.be}
\affiliation{Department of Physics, University of Antwerp,
Groenenborgerlaan 171, B-2020 Antwerpen, Belgium}
\affiliation{Key Laboratory of Materials Physics, Institute of Solid State Physics, Chinese Academy of Sciences, Hefei 230031, China}
\author{D. Moldovan}
\email{dean.moldovan@uantwerpen.be}
\affiliation{Department of Physics, University of Antwerp,
Groenenborgerlaan 171, B-2020 Antwerpen, Belgium}
\author{W. Xu}
\affiliation{Key Laboratory of Materials Physics, Institute of Solid State Physics, Chinese Academy of Sciences, Hefei 230031,
China}
\affiliation{Department of Physics, Yunnan University, Kunming 650091, China}
\author{F. M. Peeters}
\email{francois.peeters@uantwerpen.be}
\affiliation{Department of Physics, University of Antwerp,
Groenenborgerlaan 171, B-2020 Antwerpen, Belgium}

\date{\today}

\begin{abstract}
Using the tight-binding approach, we investigate the electronic properties of bilayer phosphorene (BLP) quantum dots (QDs) in the presence of perpendicular electric and magnetic fields. Since BLP consists of two coupled phosphorene layers, it is of interest to examine the layer-dependent electronic properties of BLP QDs, such as the electronic distributions over the two layers and the so-produced layer-polarization features, and to see how these properties are affected by the magnetic field and the bias potential. We find that in the absence of a bias potential only edge states are layer-polarized while the bulk states are not, and the layer-polarization degree (LPD) of the unbiased edge states increases with increasing magnetic field. However, in the presence of a bias potential both the edge and bulk states are layer-polarized, and the LPD of the bulk (edge) states depends strongly (weakly) on the interplay of the bias potential and the interlayer coupling. At high magnetic fields, applying a bias potential renders the bulk electrons in a BLP QD to be mainly distributed over the top or bottom layer, resulting in layer-polarized bulk Landau levels (LLs). In the presence of a large bias potential that can drive a semiconductor-to-semimetal transition in BLP, these bulk LLs exhibit different magnetic-field dependences, i.e., the zeroth LLs exhibit a linear-like dependence on the magnetic field while the other LLs exhibit a square-root-like dependence.
\end{abstract}

\maketitle

\section{Introduction}

Two-dimensional (2D) black phosphorus (BP) is a direct-band-gap semiconducting material, which has been recently fabricated through exfoliation methods \cite{Li2014, Liu2014, Brent2014}. Bulk BP is a layered material in which individual single layers are stacked and coupled via the van der Waals interaction \cite{Cartz1979}. A single layer of BP is called phosphorene, where each atom is covalently bonded with three neighboring atoms via $sp^3$ hybridization, thereby forming a puckered hexagonal lattice \cite{Morita1986}. Due to this unique lattice structure of phosphorene, 2D BP exhibits strongly anisotropic electronic, optical and transport properties \cite{Rodin2014,Xia2014,Qiao2014}, which are atypical for most 2D materials. Encapsulation of 2D BP with hexagonal boron nitride leads to the formation of a 2D electron gas with high electron mobility, which allows the observation of magnetic quantum oscillations \cite{Chen2015} and the quantum Hall effect \cite{Li2016}. One of the most striking characteristics of 2D BP is its strong response to external strain and bias. It was shown that the electronic properties of single-layer and bilayer BP can be tuned by applying external strain and/or bias\cite{Cakir2014,LiY2014}. In particular, an external bias can drive a semiconductor-to-semimetal transition in bilayer BP \cite{Perira2015,Yuan2016}, leading to the appearance of Dirac-like cones and parabolic-like bands (inverted) in its energy spectrum \cite{Wu2017}. In addition to strain and bias, edge effects also play an important role in affecting the physical properties of 2D BP nanostructures. For instance, armchair- and zigzag-terminated nanoribbons of 2D BP exhibit different scaling rules for the band gap versus the ribbon width ($E_g\sim1/W$ and $E_g\sim 1/W^2$, respectively, with $E_g$ the band gap and $W$ the ribbon width) \cite{Tran2014}.

Recently, BP quantum dots (QDs) have been fabricated through chemical methods \cite{ZhangX2015, Sun2015}. The obtained QDs have a lateral size of several nanometers and a vertical thickness of few layers. Therefore, one may expect significant confinement and edge effects in such small nanoscale QDs. Motivated by these experiments, theoretical studies have been carried out on the electronic and optical properties of monolayer phosphorene (MLP) QDs \cite{ZhangR2015, Niu2016, Li2017}. Interesting results were obtained, such as unconventional mid-gap edge states \cite{ZhangR2015}, anomalous size dependence of optical emission gap \cite{Niu2016}, and robust magneto-optical absorption by edge states \cite{Li2017}. In addition to MLP QDs, the electronic properties of MLP quantum rings (QRs) were also investigated recently and Aharonov-Bohm oscillations were predicted in the energy spectrum of such QRs \cite{llli2017}. Because the electronic, optical and transport properties of 2D BP are also strongly dependent on its layer number \cite{Qiao2014, Xu2015, Wang2015, ZhangG2015}, it is both of fundamental and practical interest to investigate the effect of the interlayer coupling on these physical properties. In this regard, bilayer phosphorene (BLP) is a natural candidate that can provide basic information on such an inter-layer coupling effect.

In the present work, we investigate theoretically the electronic properties of BLP QDs in the presence of perpendicular electric and magnetic fields. Within the tight-binding (TB) approach, the energy levels, wave functions, density of states, and layer-dependent electronic properties of BLP QDs are obtained numerically as a function of perpendicular magnetic field and of perpendicular bias potential. The effects of the QD size and the edge type on the electronic properties of BLP QDs are also investigated. Here for simplicity, we consider square-shaped BLP QDs with zigzag and armchair edges as our model QDs. Although realistic BLP QDs could have more complex (irregular) shapes and edges, such simple model QDs may provide basic insights into important confinement and edge effects.

The main results obtained in this work are as follows: (i) Distinctive edge and bulk states are present in BLP QDs and they exhibit different responses to perpendicular electric and magnetic fields; (ii) In small-sized BLP QDs edge states may couple with bulk states. This bulk-edge coupling is not present in large-sized BLP QDs and it decreases and eventually disappears with increasing magnetic field; (iii) Edge and bulk states in BLP QDs exhibit different layer-dependent electronic properties, such as layer-resolved electronic distributions and so-produced layer-polarization features, and these layer-dependent properties can be manipulated by perpendicular electric and magnetic fields. In addition, magneto-electronic properties of unbiased and biased BLP QDs are analysed in detail, which is essential for the understanding of other important physical properties, such as electrically tunable magneto-optical and magneto-transport properties.

This paper is organized as follows. In Sec. II, we present the TB model approach for studying the electronic properties of BLP QDs in the presence of perpendicular electric and magnetic fields. In Sec. III, we briefly investigate the effect of a bias potential on the electronic band structure of bulk BLP. In Sec. IV, the main results are presented and analysed for the magnetic-field and bias-potential dependencies of layer-dependent electronic properties (i.e., energy levels, wave functions and density of states) of BLP QDs. Finally, we make a summary and give concluding remarks in Sec. V.

\begin{figure}[htbp]
\centering
\includegraphics[width=0.35\textwidth]{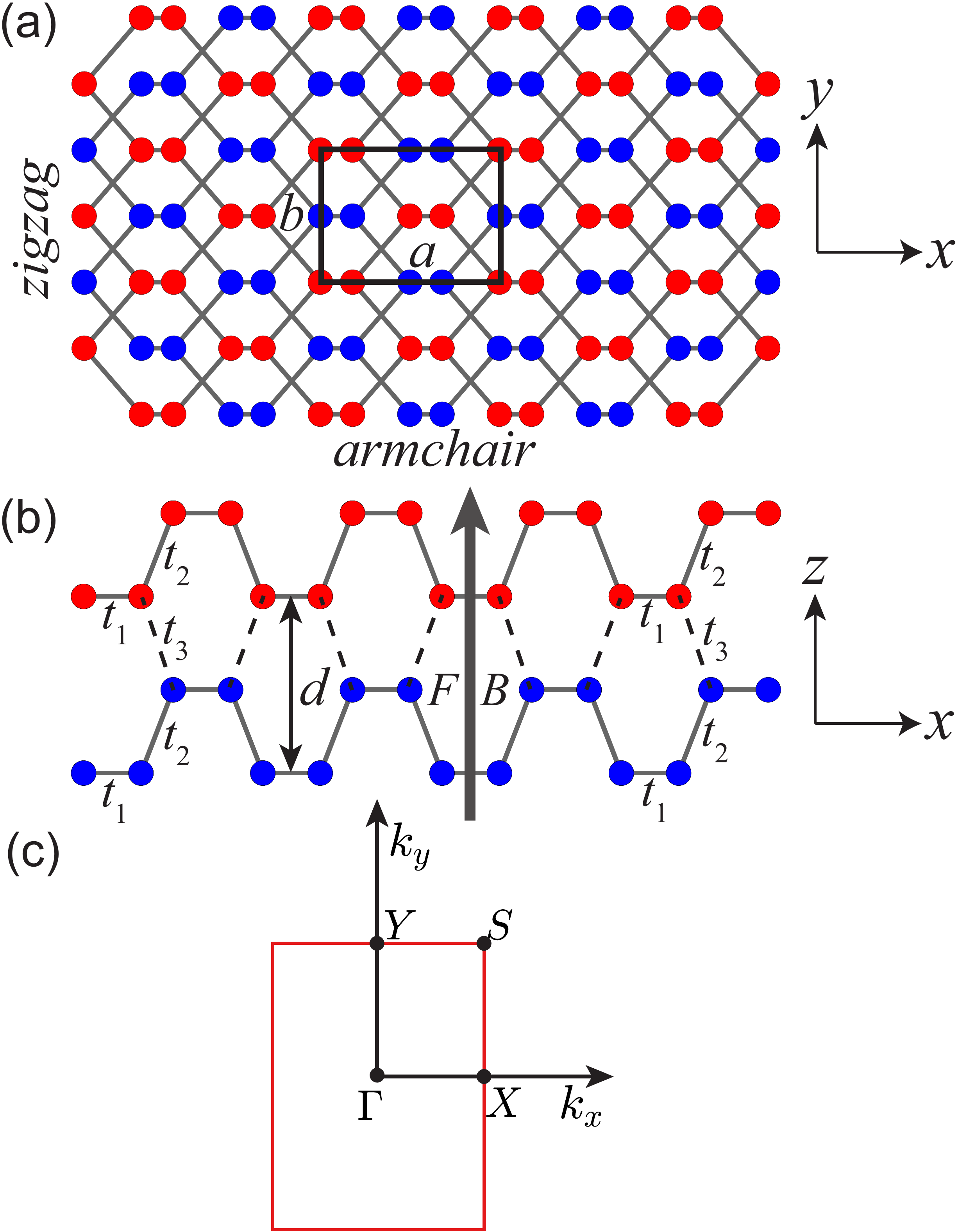}
\caption{(a) Top and (b) side views of the lattice structure of AB-stacked BLP. The upper and lower phosphorene layers, with interlayer separation $d$, are depicted by red and blue phosphorus atoms, respectively. The armchair (zigzag) edges are assumed to be along the $x$ ($y$) direction. External electric and magnetic fields, denoted by $F$ and $B$, are applied perpendicular to the phosphorene layers (i.e. along the $z$ direction). The black rectangle represents the unit cell of BLP with $a$ and $b$ being the lattice constants. The symbols $t_1$ and $t_2$ are the intralayer hopping parameters and $t_3$ is the interlayer hopping parameter. (c) Schematic plot of the first Brillouin zone (the red rectangle) with four high-symmetry points $\Gamma$, $X$, $Y$, and $S$ as indicated by black dots.} \label{fig1}
\end{figure}

\section{Theoretical Approach}
We consider AB-stacked (Bernal) BLP consisting of two phosphorene layers coupled via the van der Waals interaction, as shown in Fig. \ref{fig1}(a). This stacking configuration is energetically most stable for BLP according to first-principle calculations \cite{Cakir2015} and can be viewed as shifting the upper and lower phosphorene layers by half unit-cell length along the armchair or zigzag direction. Due to the puckered lattice structure, phosphorene has two atomic sublayers and thus BLP has four atomic sublayers, see Fig. \ref{fig1}(b). Low-energy electrons and holes in BLP are described by the following TB Hamiltonian:
\begin{equation}\label{e1}
H=\sum_{i}\varepsilon_ic_i^{\dag}c_i+\sum_{i\neq j}t_{ij}^{\|}c_i^{\dag}c_j+\sum_{i\neq j}t_{ij}^{\perp}c_i^{\dag}c_j,
\end{equation}
where the summation runs over all lattice sites of the system,
$\varepsilon_i$ is the on-site energy at site $i$, $t_{ij}^{\|}$ ($t_{ij}^{\perp}$) is the intralayer (interlayer) hopping energy between sites $i$ and $j$, and $c_i^{\dag}$ ($c_j$) is the creation (annihilation) operator of an electron at site $i$ ($j$). This TB model has been proposed for MLP and BLP \cite{Rudenko2014,Rudenko2015}, and has been shown to accurately reproduce the band structures of MLP and BLP obtained from DFT-GW calculations over a wide energy range. However, as pointed out by the previous works \cite{Rudenko2014,Ezawa2014,Jiang2015}, the main features of the band structure of MLP can be qualitatively described by a minimal TB model that only takes into account the two largest hopping parameters ($t_1$ and $t_2$). For BLP, additional hopping parameters are required to describe the interlayer coupling effect, and in a minimal TB model only the nearest-neighbor interlayer hopping parameter ($t_3$) needs to be taken into account since it crucially determines the band gap of BLP. With these hopping parameters ($t_1$, $t_2$ and $t_3$), the band gap can be evaluated as $E_g=2t_2+4t_1$ for MLP and $E_g=2t_2\sqrt{1+(t_3/t_2)^2}+4t_1-2t_3$ for BLP. From these two expressions, it can be seen that the band gap of BLP is smaller than that of MLP, which is induced by the third hopping parameter $t_3$ characterizing the interlayer coupling effect in BLP. We found that by choosing $t_1=-1.21$ eV, $t_2=3.18$ eV, and $t_3=0.22$ eV 
the band gaps of MLP and BLP calculated from this minimal TB model are given by $E_g^{\textrm{MLP}}=1.51$ eV and $E_g^{\textrm{BLP}}=1.09$ eV, which agree with those obtained by the full TB model that includes five intralayer hoppings for MLP and additional four interlayer hoppings for BLP \cite{Rudenko2014}. Moreover, the corresponding band structures of MLP and BLP near their band gaps are also found to be in agreement with those obtained by the full TB model. These motivate us to employ this three-parameter TB model to study the low-energy electronic properties of BLP.

In the presence of a perpendicular electric field, the four atomic sublayers in BLP will gain different on-site electrostatic potentials in the form of $(1/2+\xi)V$, $(1/2-\xi)V$, $(-1/2+\xi)V$, and $(-1/2-\xi)V$, where $V=eFd$ is the electrostatic potential difference between the top and bottom phosphorene layers, with $e$ the elementary charge, $F$ the electric field strength, $d$ the interlayer separation, and $\xi=0.202$ is a linear scaling factor that accounts for the sublayer dependence of the on-site electrostatic potential \cite{Yuan2016}. The effect of applying a perpendicular magnetic field to BLP is incorporate into the TB Hamiltonian (1) via the Peierls substitution, which modifies the intralayer and interlayer hopping energies as $t_{ij}^{\|,\perp} \to t_{ij}^{\|,\perp}\exp\big[i(2\pi/\Phi_0)\int_i^j\textbf{A}\cdot d\textbf{l}\big]$, where $\Phi_0=h/e$ is the magnetic flux quantum with $h$ the Planck constant, and $\textbf{A}=(0, Bx, 0)$ is the magnetic vector potential in the Landau gauge with $B$ the magnetic field strength. The magnetic flux threading a plaquette is defined as $\Phi=Bab$ in units of the flux quantum $\Phi_0$, with $a=4.37$ {\AA} and $b=3.31$ {\AA} the two in-plane lattice constants of phosphorene.

BLP QDs can be modelled by cutting an infinite BLP sheet into small-area flakes with various geometric shapes (e.g. rectangle, triangle, hexagon and circle) and with different edge types (e.g. zigzag, armchair and disordered). Here for simplicity, we consider square-shaped BLP QDs with zigzag and armchair edges as our model QDs. Although realistic BLP QDs could have more complex (irregular) shapes and edges, such simple model QDs may provide basic insights into crucial edge and confinement effects. The energy levels and wave functions of square-shaped BLP QDs subjected to perpendicular electric and magnetic fields are obtained by numerically solving the TB model. All numerical TB calculations are performed using the recently developed PYBINDING package \cite{Moldovan2016}.

With the energy levels obtained, the electronic density of states (DOS) is computed as $\textrm{DOS}(E)=\sum_{n}\exp[-(E-E_n)^2/\Delta^2]$, with $E$ the given energy, $n$ the state index, $E_n$ the energy level, and $\Gamma$ the broadening factor. In the present work, unless otherwise specified, $\Delta=5$ meV is adopted throughout the DOS calculations.

\begin{figure*}[htbp]
\begin{center}
\includegraphics[width=0.85\textwidth]{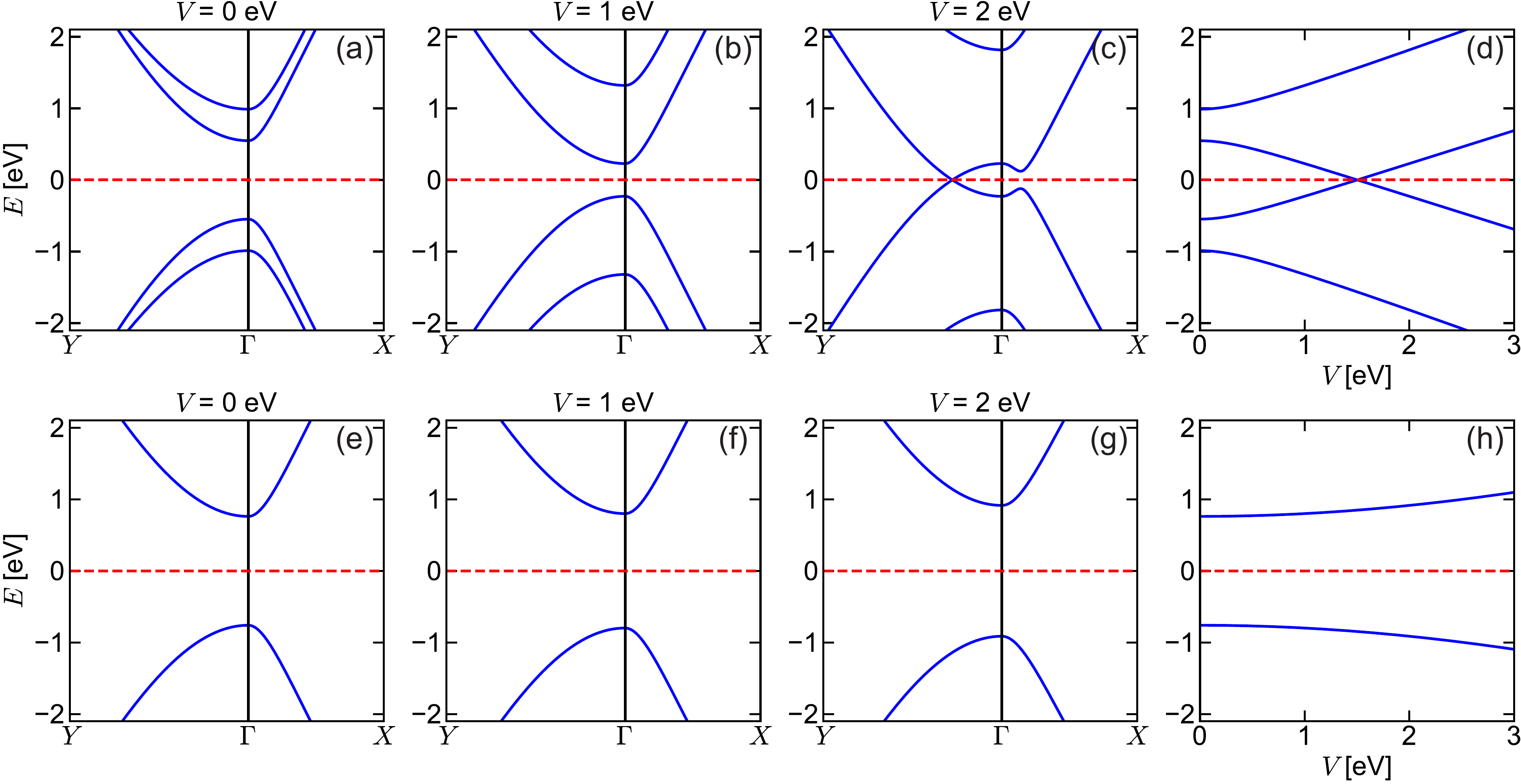}
\caption{Band structures of inifite BLP (MLP) subjected to different bias potentials $V$ as indicated in (a)-(c) [(e)-(g)] and the band energies at the $\Gamma$ point as a function of $V$ shown in (d) [(h)]. The red dashed line denotes the zero energy reference.
} \label{fig2}
\end{center}
\end{figure*}

\begin{figure*}[t]
\begin{center}
\includegraphics[width=0.92\textwidth]{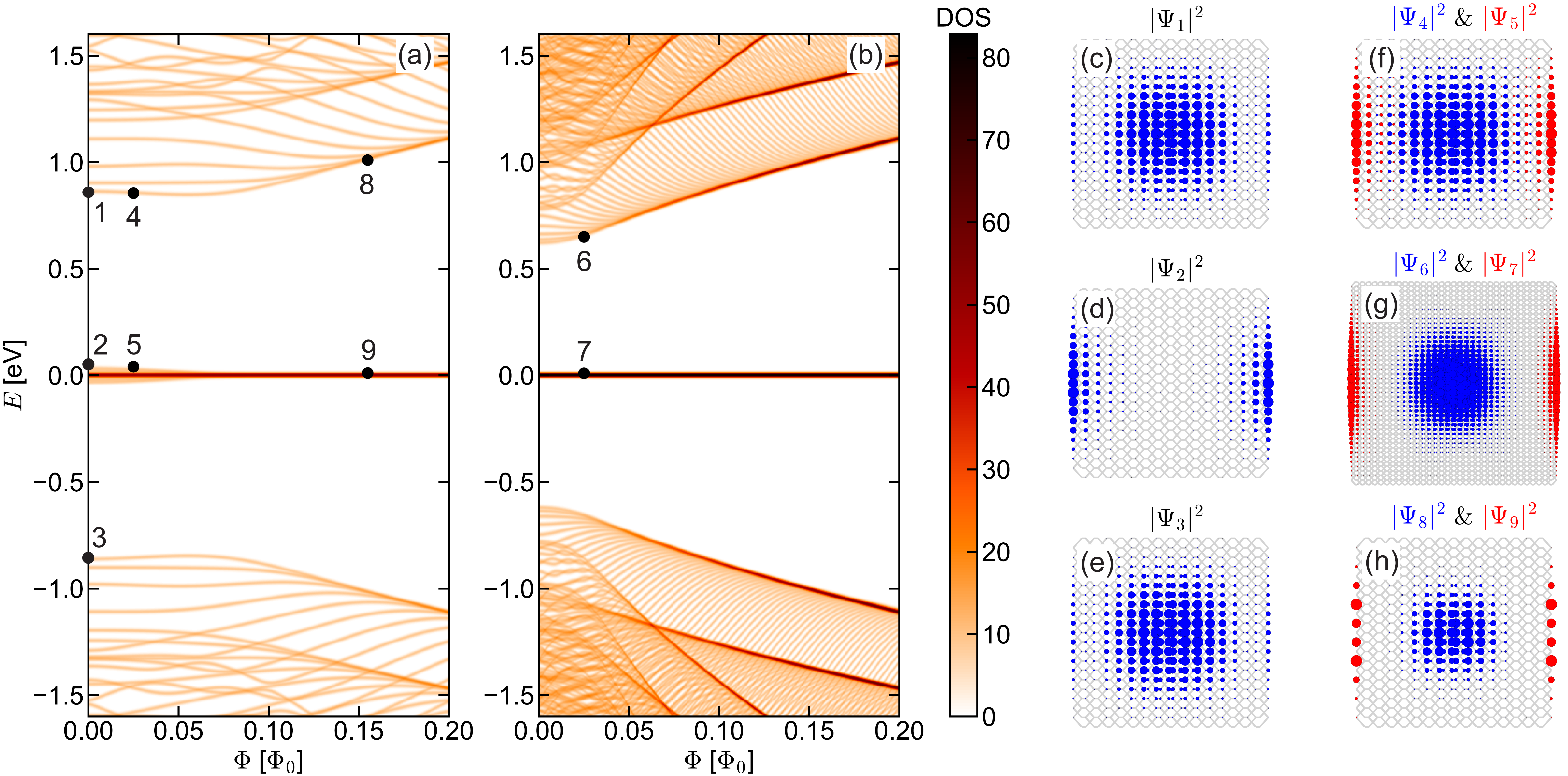}
\caption{(a) [(b)] Energy levels of an unbiased BLP QD, with dot size $L=3.5$ (7.5) nm, as a function of the magnetic flux $\Phi$, where the colorbar shows the DOS values at these levels. (c)-(e) Probability densities of the electronic states denoted by the points 1-3 shown in (a). (f)-(h) Probability densities of the electronic states denoted by the points 4-9 shown in (a) and (b). The size of the blue/red dots shown in (c)-(h) represents the amplitude of the electronic probability density.} \label{fig3}
\end{center}
\end{figure*}

\section{Bulk BLP: Bias Effect}

Before diving into BLP QDs, we first consider the effect of a perpendicular electric field (i.e. the bias effect) on the band structure of bulk BLP. Due to the in-plane translational invariance, a Fourier transform is performed to convert the real-space TB Hamiltonian \eqref{e1} into momentum space, and then the corresponding Hamiltonian is numerically diagonalized to obtain the band structure.

In Figs. \ref{fig2}(a)-\ref{fig2}(d), we show the band structure of bulk BLP for different bias potentials $V$ as indicated in (a)-(c) and the band energies at the $\Gamma$ point as a function of $V$ in (d). As can be seen, unbiased BLP has an anisotropic band structure with a finite direct band gap, which is inherently due to the puckered lattice structure of phosphorene. Applying a bias gradually reduces the band gap and eventually drives a semiconductor-to-semimetal transition. The critical bias potential for such a transition is found to be $V_c\simeq1.5$ eV, in agreement with recent theoretical work \cite{Perira2015}. In the semimetal phase, biased BLP exhibits an interesting anisotropic band structure: it is \textit{linear-like and gapless} along the $\Gamma$-$Y$ direction, but \textit{quadratic-like and inverted} along the $\Gamma$-$X$ direction. Consequently, Dirac-like cones, parabolic-like bands and band inversions coexist in biased semimetallic BLP. For comparative purposes, we also show in Figs. \ref{fig2}(e)-\ref{fig2}(h) the effect of a perpendicular electric field on the band structure and the corresponding $\Gamma$-point band energies of bulk MLP: (e)-(g) for the band structure and (h) for the band gap. As can be seen, applying a bias only increases the band gap of MLP while it has little effect on the main characteristics of the band dispersion, in sharp contrast to the case of BLP. This large difference is attributed to the interlayer coupling effect in BLP, which is also responsible for the band-gap decrease in BLP as compared to that in MLP. The electrically tunable band structure of BLP is expected to have important consequences on the electronic properties of biased BLP QDs in the presence of a perpendicular magnetic field.

\section{BLP QD: Magnetic and Bias Effects}

Now we turn to the study of the electronic properties of BLP QDs in the presence of perpendicular magnetic field and bias potential. Because the finite size breaks in-plane translational invariance, the real-space TB Hamiltonian \eqref{e1} is directly diagonalized to obtain the eigenenergies and eigenfunctions of the electronic states in BLP QDs. In the following, the effects of magnetic field and bias potential on the electronic properties of such QDs are investigated.

\subsection{Magnetic-field effect}

In Figs. \ref{fig3}(a) and \ref{fig3}(b), we show the DOS-projected energy levels of an unbiased BLP QD (i.e. $V=0$), as a function of the magnetic flux $\Phi$, for different dot sizes $L$: (a) $L=3.5$ nm and (b) $L=7.5$ nm. Because of the small size of the QD, a large magnetic field ($B = 2850$ T for $\Phi = 0.1$ $\Phi_0$) is required in order to produce a significant influence on the energy levels. Nevertheless, as the influence of the magnetic field scales with the magnetic flux threading the QD, similar results will be obtained for smaller magnetic fields if larger-sized QDs are considered.

\begin{figure*}[htbp]
\begin{center}
\includegraphics[width=0.72\textwidth]{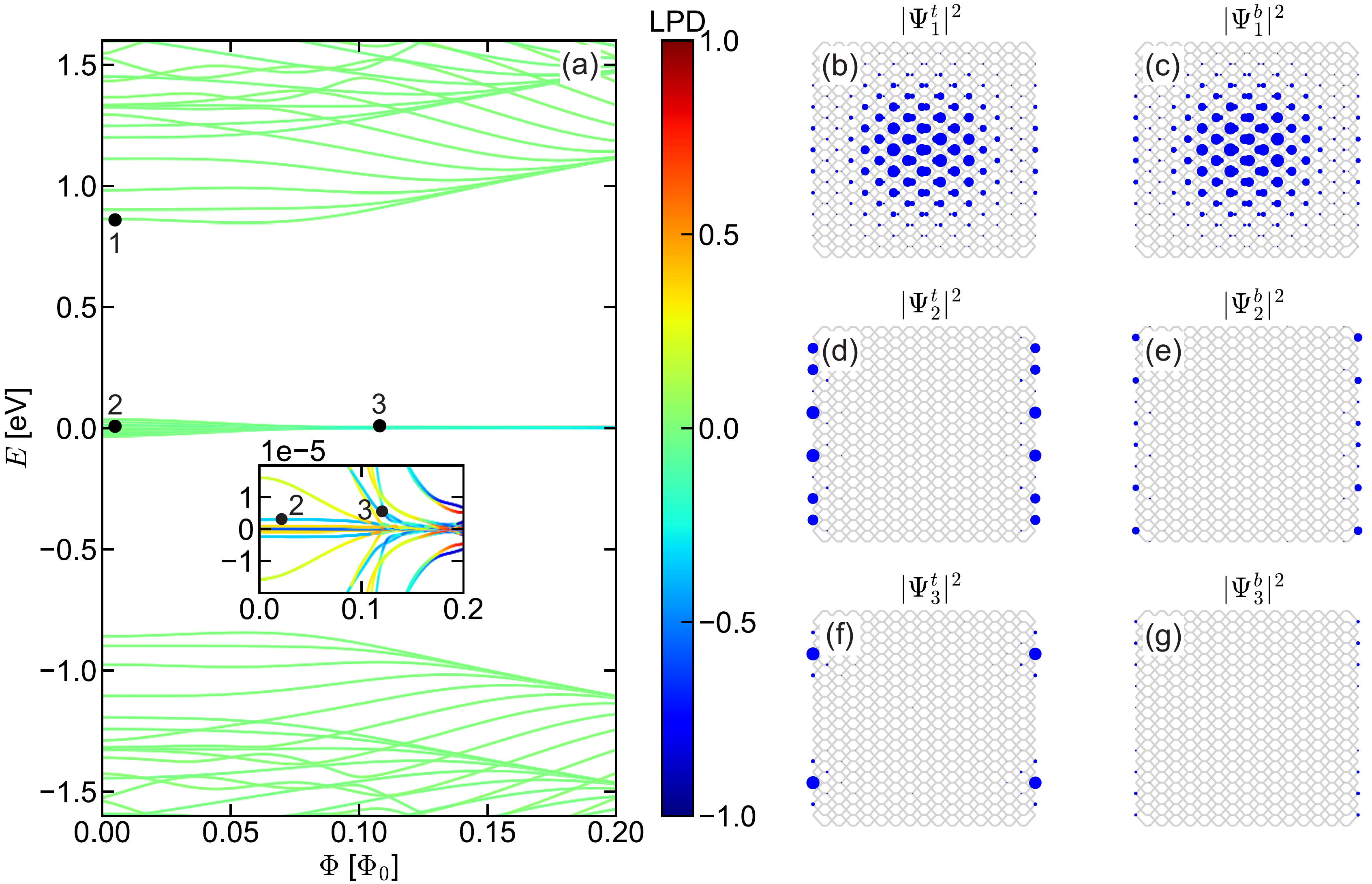}
\caption{(a) Energy levels of an unbiased BLP QD with dot size $L=3.5$ nm, as a function of the magnetic flux $\Phi$, where the colorbar shows the LPD ($\eta$) values at these levels. The inset shown in (a) shows a zoom of the edge levels around zero energy. (b)-(g) Layer-resolved probability densities of the electronic states indicated by the points 1-3 shown in (a). $|\Psi_j^t|^2$ and $|\Psi_j^b|^2$ ($j=1, 2, 3$) denote the top-layer and bottom-layer electronic distributions, respectively. The size of the blue dots shown in (b)-(g) represents the amplitude of the electronic probability density.} \label{fig4}
\end{center}
\end{figure*}

As can be seen in Figs. \ref{fig3}(a) and \ref{fig3}(b), there are nearly flat energy levels within the band gap of the BLP QD. These energy levels correspond to the edge states, while those above (below) the band gap correspond to the conduction (valence) bulk states. Here we distinguish edge and bulk states in terms of their wave-function properties: the former are strongly localized at the QD boundary while the latter are mainly distributed around the QD centre, see Figs. \ref{fig3}(c)-\ref{fig3}(e), which show the probability densities $|\Psi|^2$ of the electronic states indicated by the points 1-3 in Fig. \ref{fig3}(a). An important feature of the edge states is that they are almost unaffected by the magnetic field, as reflected by the quasi-flat energy levels shown in Figs. \ref{fig3}(a) and \ref{fig3}(b). This feature is attributed to the strong localized nature of the edge states. Furthermore, we find that the DOS of the edge states is larger than that of the bulk states. This is because the edge levels are spaced very closely to each other and thus they can be viewed as nearly degenerate.

However, when comparing Figs. \ref{fig3}(a) and \ref{fig3}(b), the following differences can be observed: (i) The smaller-sized QD has a larger band gap and also a larger energy-level separation due to the stronger confinement effect; (ii) The larger-sized QD has a larger electronic DOS for both the bulk and edge states. This is because the bulk (edge) DOS is proportional to the number of atoms in the QD centre (at the zigzag boundary) and thus both of them increase with the dot size; (iii) At high magnetic fields, the bulk levels in the larger-sized QD converge to distinct Landau levels (LLs) with a large DOS due to the high LL degeneracy; (iv) The edge states in the smaller-sized QD are less unaffected by the magnetic field and have a certain band-like broadening at lower magnetic fields. This broadening comes from the bulk-edge coupling in the smaller-sized QD, while it is absent in the larger-sized QD where the edge states have no such band-like broadening.

To see clearly the presence (absence) of the bulk-edge coupling in the smaller-sized (larger-sized) QD, we show in Figs. \ref{fig3}(f) and \ref{fig3}(g) the probability densities of the bulk and edge states denoted by the points 4, 5 in Fig. \ref{fig3}(a) and those denoted by the points 6, 7 in Fig. \ref{fig3}(b). We also find that the bulk-edge coupling in the smaller-sized QD becomes much weaker at high magnetic fields, see Fig. \ref{fig3}(h), which shows the probability densities of the bulk and edge states denoted by the points 8, 9 in Fig. \ref{fig3}(a). This is because the bulk states are more confined around the QD centre at high magnetic fields due to the strong magnetic confinement.

Since BLP is made of two coupled phosphorene layers with each one having two atomic sublayers, it is natural to think of studying the electronic-state distributions over these two atomic layers or four atomic sublayers. In doing so, we need to look into the layer-resolved electronic probability densities for both the bulk and edge states. Therefore, we define a new physical quantity $\eta$, the layer-polarization degree (LPD), for the electronic states in a BLP QD, which characterizes how much an electronic state is distributed over the top and bottom layers, and mathematically, this quantity is given by
\begin{equation}\label{e2}
\eta=\frac{\int|\Psi^t(r)|^2dr-\int|\Psi^b(r)|^2dr}
{\int|\Psi^t(r)|^2dr+\int|\Psi^b(r)|^2dr},
\end{equation}
where $\Psi^t(r)$ and $\Psi^b(r)$ are the electronic probability densities in the top and bottom layers, respectively. Thus, $\eta=1$ ($-1$) indicates that the electronic states are only distributed over the top (bottom) layer (i.e. completely layer-polarized), $\eta=0$ that the electronic states are symmetrically distributed over the two layers (i.e. fully layer-unpolarized), and $0<|\eta|<1$ that the electronic states are asymmetrically distributed over the two layers (i.e. partially layer-polarized).

\begin{figure*}[t]
\begin{center}
\includegraphics[width=0.83\textwidth]{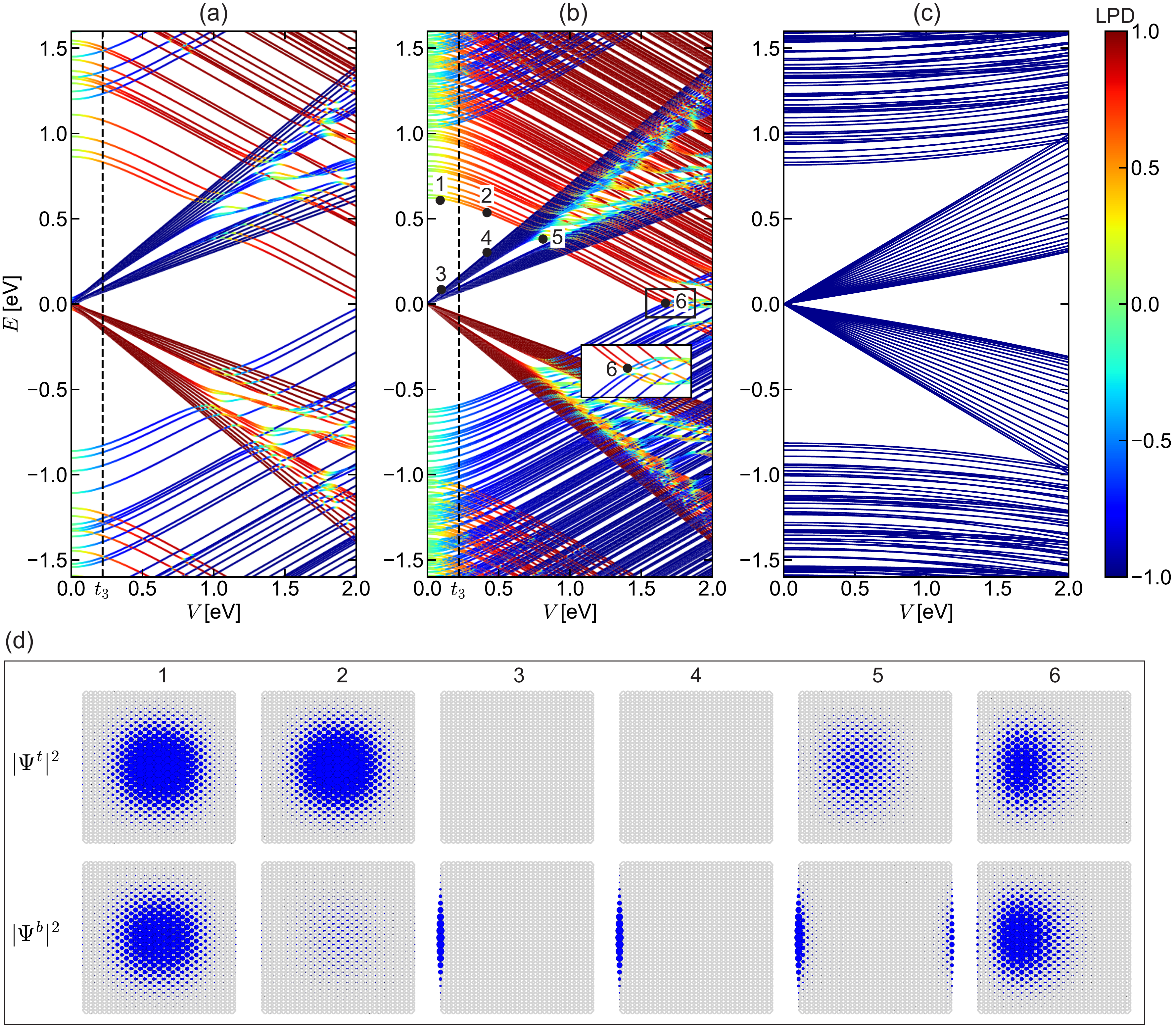}
\caption{(a) [(b)] LPD-projected energy levels of a BLP QD with dot size $L=3.5$ (7.5) nm, at zero magnetic field ($\Phi=0$), as a function of the bias potential $V$. (c) Results of a MLP QD with dot size $L=7.5$ nm shown for comparative purposes. (d) Layer-resolved probability densities $|\Psi^t|^2$ and $|\Psi^b|^2$ of the electronic states denoted by the points 1-6 shown in (b). Here, the black rectangle shown in (b) is enlarged as an inset, and the vertical dashed lines shown in (a) and (b) mark the position at which the bias potential $V$ is equal to the interlayer coupling energy $t_3$.} \label{fig5}
\end{center}
\end{figure*}

In Fig. \ref{fig4}(a), we show the LPD-projected energy levels of the same BLP QD as in Fig. \ref{fig3}(a). The inset plot shows more clearly the edge-state levels that are very close to each other around zero energy. As can be seen, the edge states inside the band gap of the QD are partially layer-polarized ($0<|\eta|<1$) while the conduction and valence bulk states outside the band gap are fully layer-unpolarized ($\eta=0$). To clearly see such layer-polarized features of the bulk and edge states, we show in Figs. \ref{fig4}(b)-\ref{fig4}(d) the layer-resolved electronic probability densities $|\Psi^t|^2$ and $|\Psi^b|^2$ of the bulk and edge states corresponding to the points 1-3 marked in Fig. \ref{fig4}(a). Indeed, those edge states are partially layer-polarized while those conduction and valence bulk states are fully layer-unpolarized. Such a difference is mainly attributed to the different interlayer symmetries for the edge and bulk states, which can be understood as follows: When cutting the BLP sheet perpendicularly into squared-shaped QDs, we find that the presence of interlayer hoppings breaks the edge symmetry between the top and bottom phosphorene layers while it keeps the bulk symmetry between the two layers, as shown in Fig. \ref{fig1}(b). Notice that the edge atom of one layer is two hops away from the interlayer hopping $t_3$ while the other layer is only one hop away from it. Because edge states are strongly localized at the zigzag boundaries, the broken edge symmetry gives rise to zero-field layer polarization of the edge states. However, due to the bulk symmetry between the top and bottom phosphorene layers, the bulk states are symmetrically distributed over the two layers and so there is no such layer polarization for the bulk states.

At lower magnetic fields, the edge states are less polarized due to their coupling to the unpolarized bulk states. With increasing magnetic field, they become more polarized because they are gradually decoupled from the bulk states (which become more localized around the QD center at high magnetic fields). Our numerical calculations also indicate a similar result for the larger-size QD, i.e., the edge (bulk) states are asymmetrically (symmetrically) distributed over the two layers and thus they are partially layer-polarized (fully layer-unpolarized). However, the difference is that the LPD values of the edge levels in the larger-size QD is almost unaffected by the magnetic field due to the absence of the bulk-edge coupling. In addition, we find that the zero-field layer polarization of edge states increases with decreasing QD size. This is because the importance of dot boundaries are more significant in smaller-size QDs due to the increased edge-to-volume ratio and the zigzag boundaries of the top and bottom layer are also more asymmetric.

\subsection{Bias effect}

Now we consider the effect of applying a perpendicular electric field (i.e. a bias potential) on the electronic properties of a BLP QD. Here, the applied bias potential $V$ is antisymmetric with respect to the $z=0$ plane, i.e. $V(z)=-V(-z)$, due to the mirror symmetry between the two layers in a BLP QD. In Figs. \ref{fig5}(a) and \ref{fig5}(b), we show the LPD-projected energy levels of a BLP QD at zero magnetic field ($\Phi=0$), as a function of the bias potential $V$, for different dot sizes $L$: (a) $L=3.5$ nm and (b) $L=7.5$ nm. For comparative purposes, we also show in Fig. \ref{fig5}(c) the results of a MLP QD with dot size $L=7.5$ nm.

As can be seen in Figs. \ref{fig5}(a) and \ref{fig5}(b), for smaller bias potentials the energy levels of the edge (bulk) states exhibit a linear (quadratic) Stark shift, while for larger bias potentials both the bulk and edge levels exhibit a linear-like Stark shift. We explain in the following the different behaviors of these Stark shifts observed for smaller and larger bias potentials.

The linear (quadratic) Stark effect exhibited by the edge (bulk) states observed for smaller bias potentials can be understood qualitatively within the framework of perturbation theory. As already shown in Figs. \ref{fig3}(a) and \ref{fig3}(b), the energy levels of the edge states are spaced very closely while those of the bulk states are not. Therefore, degenerate (non-degenerate) perturbation theory can be used to study the response of the edge (bulk) states to the external bias. For simplicity, we use a two-level model to explain the linear (quadratic) Stark effect observed for the edge (bulk) states in the BLP QD. Here for both the edge and bulk states, the two unperturbed (unbiased) energy levels are denoted by their eigenenergies $E_1$, $E_2$ and corresponding eigenfunctions $\ket{1}$, $\ket{2}$, and for the edge states $E_1\simeq E_2$ can be reasonably assumed due to their nearly degenerate energy levels. We further denote the edge-state or bulk-state Hamiltonian in the presence of an external bias as $H=H_0+V(z)$ with $H_0$ the unperturbed Hamiltonian and $V(z)$ the layer-dependent bias potential. For the unperturbed Hamiltonian $H_0$, we have $H_0\ket{j}=E_{j}\ket{j}$ ($j=1$, $2$) and we apply perturbation theory with respect to the bias potential $V(z)$.

Within non-degenerate perturbation theory, the energy of the perturbed bulk states can be obtained up to second order as
\begin{equation}\label{e3}
E_j=E_j + \bra{j}V(z)\ket{j} + \frac{|\bra{j}V(z)\ket{i}|^2}{E_j-E_i},
\end{equation}
with $j=1$, $2$ and $i\neq j$. As mentioned previously, the unbiased bulk states have symmetric electronic distributions over the two layers, i.e., their wave functions are symmetric with respect to the $z=0$ plane. However, the perturbation (bias) potential is anti-symmetric with respect to the $z=0$ plane, namely $V(z)=-V(-z)$. Therefore, the first-order term in Eq. \eqref{e3} is zero while the second-order one is non-zero. This simple result may qualitatively explain the quadratic Stark effect observed for the bulk states in the lower bias-potential region. The perturbed edge-state Hamiltonian matrix $H$ can be obtained within first-order degenerate perturbation theory, in a basis set composed of unperturbed eigenfunctions $\{\ket{1}, \ket{2}\}$ as
\begin{equation}\label{e4}
[H]=\left[\begin{array}{cc}
E_0 \ \ &  \Lambda \\
\Lambda^{\dag} \ \ &  E_0 \\
\end{array}\right],
\end{equation}
where $E_0=E_1=E_2$ and $\Lambda=\bra{1}V(z)\ket{2}$. Diagonalizing this Hamiltonian matrix, we obtain the energy levels of the edge states under the bias perturbation as $E_{\pm}=E_0\pm|\Lambda|$, which may qualitatively explain the linear Stark effect observed for the edge states in the lower bias-potential region.

However, for larger bias potentials, perturbation theory is no longer valid and can not be applied to explain the observed linear-like Stark shifts for both the bulk and edge states in the BLP QD. In this case, one has to resort to the TB Hamiltonian itself [Eq. \eqref{e1}], where the bias potential is included as the on-site term. By applying a bias potential $V$, both the interior and boundary atoms in the top (bottom) layer gain the same on-site potential $+V/2$ ($-V/2$). That is why both the bulk and edge states in the BLP QD exhibit a linear-like Stark effect at large bias potential. Furthermore, because a large bias potential makes both states layer-polarized, all the top-layer (bottom-layer) states feel only the on-site potential $+V/2$ ($-V/2$) and thus their energy levels move up (down) with $V$, see the blue and red curves shown in Figs. \ref{fig5}(a) and \ref{fig5}(b). However, this is not the case for the bulk and edge states in the MLP QD due to the absence of the layer-polarization feature. Because all the bulk/edge states are distributed only over one layer and they feel the same on-site potential $+V/2$ (or $-V/2$), their energy levels move up (or down) with increasing $V$, as shown in Fig. \ref{fig5}(c).

It is worth noting that in Figs. \ref{fig5}(a) and \ref{fig5}(b) the energy levels of the edge states in the biased BLP QD are split into four branches. However, the energy levels of the edge states in the biased MLP QD are only split into two such branches, see Fig. \ref{fig5}(c). This difference is attributed to the interlayer coupling effect. The four branches of edge states in the BLP QD have opposite linear Stark responses to the external bias: two of them go up while the rest two go down, as the bias potential increases. Likewise, the conduction and valence bulk states in the BLP QD have opposite quadratic Stark responses to the external bias. For instance, the energy of the lowest (highest) conduction (valence) bulk states decreases (increases) with the bias potential, thereby leading to a decrease in the band gap of the BLP QD. Therefore, those two branches of edge states that go up (down) with increasing bias potential eventually merge into the conduction (valence) bulk states at large bias potentials. Notice that this merging may lead to anticrossings between the energy levels of those bulk and edge states, see Fig. \ref{fig5}(a). These anticrossings are more pronounced for the smaller BLP QD because of the larger bulk-edge coupling as mentioned above. The energy levels before or after anticrossings correspond to the pure bulk and edge states, while at anticrossings they corresponds to the mixed bulk/edge states.

It is also of interest to look at the LPD ($\eta$) values of the energy levels of the bulk and edge states shown in Fig. \ref{fig5}(b). From the line colors that characterize the $\eta$ values, we observe the following features: (i) The bulk states are weakly layer-polarized when the bias potential is less than the interlayer coupling energy (i.e. $V<t_3$); (ii) With increasing bias potential such that $V>t_3$, the bulk states become strongly layer-polarized; (iii) The edge states are always layer-polarized no matter $V<t_3$ or $V>t_3$; (iv) Once the edge states merge into the bulk states, their mixed states become less layer-polarized as compared to the unmixed bulk and edge states; (v) At large bias potentials, the lowest conduction bulk states are coupled to the highest valence bulk states, leading to anticrossings between their energy levels, see the inset shown in Fig. \ref{fig5}(b). Although both the uncoupled conduction and valence bulk states are layer-polarized, the coupled conduction-valence bulk states are layer-unpolarized due to the electron-hole symmetry of the energy spectrum.

To see the above features more clearly, we choose typical bulk and edge states marked by the points 1-6 shown in Fig. \ref{fig5}(b) and plot their layer-resolved probability densities in Fig. \ref{fig5}(d). As can be seen, the bulk state corresponding to the point 1 has almost equal electronic distribution over the top and bottom layers, implying that the bulk states are almost layer-unpolarized when $V<t_3$, which is due to the dominant interlayer coupling effect. When $V>t_3$, the bulk states become layer-polarized because of the stronger bias effect, see the result corresponding to point 2. However, the biased edge states are always layer-polarized no matter $V>t_3$ or $V>t_3$, which is almost unaffected by the bias potential, see the results corresponding to points 3 and 4. Once the edge states merge into the bulk states, their mixed states have finite electronic distributions in both the top and bottom layers, see the result corresponding to point 5, implying that the mixed states are less layer-polarized as compared to the unmixed individual states. At large bias potentials, the layer-polarized conduction bulk states are coupled to the layer-polarized valence bulk states. The coupled conduction-valence bulk states have almost equal electronic distribution over the two layers, see the result corresponding to point 6, implying that they are layer-unpolarized.

Additionally, when comparing Figs. \ref{fig5}(b) and \ref{fig5}(c), we find that the band gap of the BLP (MLP) QD decreases (increases) with increasing bias potential. This is a consequence of the different bias-potential dependences of their bulk counterparts (i.e. infinite BLP and MLP), see Fig. \ref{fig2}. However, unlike infinite BLP shown in Fig. \ref{fig2}, the band gap of the BLP QD can not be closed completely with increasing bias potential, see the inset shown in Fig. \ref{fig5}(b). This difference is clearly induced by the finite-size effect in the BLP QD.

\subsection{Magnetic-field effect in the presence of bias}

\begin{figure*}[t]
\begin{center}
\includegraphics[width=0.87\textwidth]{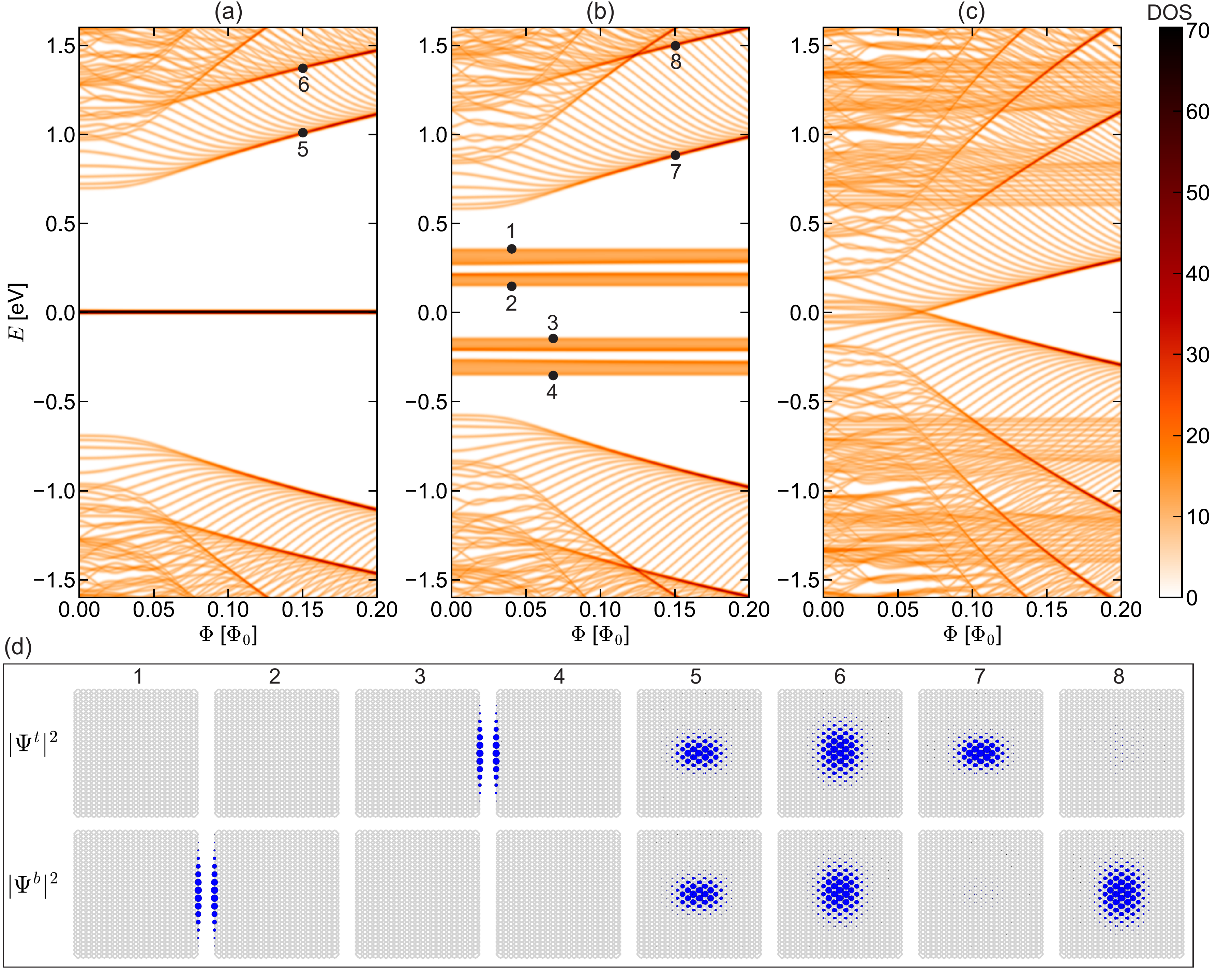}
\caption{DOS-projected energy levels of a BLP QD with dot size $L=5.5$ nm, as a function of the magnetic flux $\Phi$, for different bias potentials $V$: (a) $V=0$ eV, (b) $V=0.5$, and (c) $V=2$ eV. (d) Layer-resolved probability densities $|\Psi^t|^2$ and $|\Psi^b|^2$ of the electronic states denoted by the points 1-8 shown in (a) and (b). The size of the blue dots shown in (d) represents the amplitude of the electronic probability density.} \label{fig6}
\end{center}
\end{figure*}

In Figs. \ref{fig6}(a)-\ref{fig6}(c), we show the DOS-projected energy levels of a BLP QD with dot size $L=5.5$ nm, as a function of the magnetic flux $\Phi$, for different bias potentials $V$: (a) $V=0$ eV, (b) $V=0.5$ eV, and (c) $V=2$ eV. As can be seen, in the presence of a finite bias potential, the single unbiased edge band is split into four individual bands and the corresponding biased edge states are not only layer-resolved but also boundary-resolved. To show this, we choose four typical biased edge states indicated by the points 1-4 shown in Fig. \ref{fig6}(b) and plot their corresponding probability densities over the two layers in Fig. \ref{fig6}(d). Clearly, these four biased edge states are localized at the different zigzag boundaries and are distributed over the different layers. Their layer- and boundary-resolved features are induced by the combined effects of perpendicular electric field and puckered phosphorene lattice, introducing electrostatic on-site potentials in a biased BLP QD that are not only layer-dependent but also sublayer-dependent.

With increasing bias potential from $V=0$ to $V=0.5$ eV, the bulk LLs at high magnetic fields become layer-polarized, see Fig. \ref{fig6}(d) corresponding to the points 5-8 shown in Figs. \ref{fig6}(a) and \ref{fig6}(b). The biased bulk-LL states marked by the points 7, 8 shown in Fig. \ref{fig6}(b) have very different layer-resolved characteristics: one of them (the other) is only distributed over the top (bottom) layer. This feature indicates that there exist two groups of biased bulk LLs with distinctive electronic distributions over the two layers. When looking up the magnetic-field dependence of the bulk LLs formed at large magnetic fields shown in Figs. \ref{fig6}(a) and \ref{fig6}(b), we find that both the unbiased and biased ones exhibit a linear-like dependence characteristic of conventional 2D electronic systems with parabolic-like energy bands.

With further increasing bias potential to $V=2$ eV, bulk BLP becomes a Dirac-like semimetal, as shown in Fig. \ref{fig2}(c). This Dirac-like semimetal phase of biased BLP features a gapless linear-like band dispersion along the $\Gamma$-$Y$ direction, a gapped parabolic-like energy spectrum along the $\Gamma$-$X$ direction, and an inverted band gap at the $\Gamma$ point. However, the confinement effect in the BLP QD opens a finite band gap at zero magnetic field ($\Phi=0$), see Fig. \ref{fig6}(c). As the magnetic field $\Phi$ increases, this band gap is first closed, leading to a semiconductor-to-semimetal transition, similar to that observed in Dirac material systems such as graphene QDs \cite{Guclu2013}; and then it is re-opened, leading to a semimetal-to-semiconductor transition, similar to that observed in semiconductor material systems with parabolic-like band inversions such as InAs/GaSb broken-gap quantum wells \cite{Chao2004}. Such two successive phase transitions induced by the magnetic field arise due to the coexistence of the gapless Dirac-like spectrum along the zigzag direction and the inverted parabolic-like spectrum along the armchair direction in biased BLP, as shown in Fig. \ref{fig2}(c). At higher magnetic fields, a significantly large band gap can be observed. We note that inside this band gap no edge states are present because they merge into the bulk states outside the band gap. Moreover, we find in the presence of a large bias potential $V=2$ eV and for larger magnetic fields, the zeroth bulk LLs exhibit a linear-like dependence on the magnetic field while the others exhibit a square-root-like dependence on the magnetic field, see Fig. \ref{fig6}(c). This feature is remarkably different than that shown in Fig. \ref{fig6}(a) or \ref{fig6}(b), which is also attributed to the coexistence of the gapless Dirac-like spectrum and the inverted parabolic-like spectrum in biased BLP.

\section{Concluding Remarks}

By means of the TB approach, we have investigated the electronic properties of BLP QDs in the presence of perpendicular electric and magnetic fields. The energy levels, wave functions, and density of states of BLP QDs were obtained as a function of magnetic field and of bias potential. We find in-gap edge states that are well separated from gapped bulk states. The edge states are strongly localized at the zigzag boundaries of the QD and as a result, they are almost unaffected by the magnetic field; while the bulk states are mainly distributed around the centre part of the QD and thus they are strongly affected by the magnetic field, resulting in distinct LLs  at high magnetic fields. However, both the edge and bulk states are found to be strongly influenced by the bias potential. For instance, their energy levels exhibit remarkably linear and quadratic Stark shifts, respectively. The different Stark effects exhibited by the edge and bulk states are qualitatively explained by using perturbation theory.

The size effect on the bulk and edge states in BLP QDs was also investigated. We found that in smaller-sized BLP QDs the edge states couple to the bulk states. Such a bulk-edge coupling decreases and eventually disappears with increasing magnetic field, because the bulk states become more confined due to the strong magnetic confinement while the edge states are almost unaffected by the magnetic field.

Since BLP is composed of two coupled phosphorene layers, the bulk and edge states in BLP QDs show interesting layer-dependent electronic properties, such as layer-resolved electronic distributions and their layer-polarization features, in the absence and presence of perpendicular electric and magnetic fields. We find that in the absence of a bias potential only edge states are layer-polarized while the bulk states are not, and the layer-polarization degree of the edge states in smaller-sized QDs increases with increasing magnetic field. However, in the presence of a bias potential both the edge and bulk states are layer-polarized, and the layer-polarization degrees of the bulk (edge) states depend strongly (weakly) on the interplay of the bias potential and the interlayer coupling. The layer-polarization features of the edge and bulk states are clearly demonstrated by their layer-resolved electronic distributions. At high magnetic fields, the applied bias renders the bulk electrons in a BLP QD to perform cyclotron motion mainly in the bottom or top layer, leading to layer-polarized bulk LLs, and consequently there are two groups of biased bulk LLs with distinctive layer-resolved electronic distributions.

We also found that in the presence of a large bias potential, semiconducting bulk BLP becomes a Dirac-like semimetal with a parabolic-like band inversion. As a consequence, with increasing magnetic field, the band gap of the BLP QD is first closed, leading to a semiconductor-to-semimetal transition, similar to that observed in Dirac material systems such as graphene QDs, and then it is re-opened, leading to a semimetal-to-semiconductor transition, similar to that observed in semiconductor material systems with parabolic-like band inversions such as InAs/GaSb broken-gap quantum wells. Moreover, due to the coexistence of the gapless Dirac-like spectrum and the inverted parabolic-like spectrum in biased BLP, at large magnetic fields the zeroth bulk LLs in biased BLP QDs exhibit a linear-like dependence on magnetic field while the other LLs exhibit a square-root-like dependence.

\section*{Acknowledgments}

This work was financially supported by the Flemish Science Foundation (FWO-Vl), the National Natural Science Foundation of China (Grant No. 11574319), and by the Chinese Academy of Sciences (CAS).

\end{document}